\documentclass[11pt]{article}
\usepackage{a4}
\usepackage[dvips]{epsfig}
\usepackage{latexsym}
\usepackage{graphicx}
\usepackage{amsmath,amssymb}
\usepackage{psfrag}
\def\setb@se#1{\baselineskip=#1 \normalbaselineskip=#1}
\setb@se{14pt}
\def\setb@se#1{\baselineskip=#1 \normalbaselineskip=#1}
\setb@se{14pt}
\newcommand{\be}{\begin{equation}}
\newcommand{\ee}{\end{equation}}
\newcommand{\beqn}{\begin{eqnarray}}
\newcommand{\eeqn}{\end{eqnarray}}
\newcommand{\bsub}{\begin{subeqnarray}}
\newcommand{\esub}{\end{subeqnarray}}
\newcommand{\disp}{\displaystyle}
\newcommand{\dr}{\partial}
\newcommand{\scr}{\scriptscriptstyle}
\def\l{\left}
\def\r{\right}
\newcounter{subequation}[equation]

\makeatletter \expandafter\let\expandafter\reset@font\csname
reset@font\endcsname
\newenvironment{subeqnarray}
  {\arraycolsep1pt
    \def\@eqnnum\stepcounter##1{\stepcounter{subequation}{\reset@font\rm
      (\theequation\alph{subequation})}}\eqnarray}%
  {\endeqnarray\stepcounter{equation}}
\makeatother

\begin{document}
\begin{titlepage}
\hbox to\hsize{%

  \vbox{%
        \hbox{LMPT/YMH}%
        \hbox{\today}%
       }}

\vspace{3 cm}

\begin{center}
\Large{Numerical and asymptotic analysis of the\\
       't Hooft-Polyakov magnetic monopole}

\vskip10mm
\large
{P.~Forg\'acs, N.~Obadia${}^{*}$
and S.~Reuillon}

\vspace{1 cm}

{\small\sl
Laboratoire de Math\'emathiques et Physique Th\'eorique\\
CNRS UMR6083\\
Universit\'e de Tours,
Parc de Grandmont\\
37200 Tours, France\\}
 \vspace{0.3 cm}
${}^{*}${\small\sl
Department of Condensed Matter Physics\\
The Weizmann Institute of Science\\
P.O. Box 26, Rehovot 76100, Israel\\}
\end{center}

\vspace{20 mm}
\begingroup \addtolength{\leftskip}{1cm} \addtolength{\rightskip}{1cm}

\begin{abstract}

A high precision numerical analysis of the static, spherically symmetric
$SU(2)$ magnetic monopole equations is carried out.
Using multi-shooting and multi-domain spectral methods,
the mass of the monopole is obtained rather precisely as a function of
$\beta=M_H/M_W$ for a large $\beta$-interval
($M_H$ and $M_W$
denote the mass of the Higgs and gauge field respectively).
The numerical results necessitated the reexamination and subsequent correction of a
previous asymptotic analysis of the monopole mass in the literature for $\beta\ll1$.

\end{abstract}

\endgroup
\end{titlepage}
\newpage

\section{Introduction}
It has been found some time ago by 't Hooft and Polyakov \cite{tHooft74,Polyak74}
that spontaneously
broken gauge theories (with a simple gauge group) admit classical solutions
with quantized magnetic charge and finite energy.
The self-energy (or mass) of these magnetic monopole solutions is well localized
at any instant of time, hence they can be interpreted as `particles'.
Furthermore these magnetic monopoles are stabilized by a quantum number of topological origin,
corresponding to their magnetic charge. Magnetic monopoles provide a natural explanation
of the quantization of the electric charge, and they
inevitably arise in grand unified theories. They have been very extensively studied,
see the recent monograph of Manton and Sutcliffe \cite{Mant-Sutcl04}
for a very detailed description of the zoo of topological solitons,
including magnetic monopoles and a bibliography.

An interesting problem for magnetic monopoles (or more generally for
particle-like solutions of finite energy) is a quantitative description
of their internal structure, their size and their mass in function of the
parameters of the underlying theory. Since the equations describing such objects are
nonlinear, simple analytic descriptions are rarely available and in most
cases one has
to resort to numerical techniques (often combined with some analytic methods).
The simplest prototype of the 't Hooft-Polyakov monopole arises in an SU(2)
gauge theory coupled to a Higgs scalar in the adjoint representation.
There are two mass scales in this model, the mass of the vector mesons, $M_{\rm W}$,
and the mass of the Higgs particle, $M_{\rm H}$. The
monopole solution of 't Hooft and Polyakov is static, spherically symmetric and it depends
on a single dimensionless parameter, $\beta=M_{\rm H}/M_{\rm W}$.
An analytic solution is available in the special case
$\beta=0$ \cite{Bogomolnyi76,Prasad-Somm75},
for non-zero values of $\beta$ monopole solutions have been studied mostly numerically
 e.g.\ \cite{tHooft74,Bogo-Marinov, Bais-Primack} and \cite{Mant-Sutcl04}. In the two limiting
cases, $\beta\ll1$, and $\beta\gg1$,
a combination of asymptotic and numerical techniques \cite{Kirk-Zachos81,Gardner82} have
permitted to derive asymptotic expansions for the mass of the monopole
as a function of the parameter $\beta$.

The aim of this paper is to adapt and test numerical techniques
based on combining standard Runge-Kutta integration with multi-domain spectral
methods, which latter have been successfully applied for 3 dimensional elliptic
problems in Numerical Relativity \cite{lorene}.
Our results demonstrate that one can achieve
very good numerical precision for a very wide range
of the parameter $\beta$, $10^{-4}\leq\beta\leq2\times10^{3}$,
and we tabulate the values of the pertinent parameters
of the solutions to 11 significant digits.
By comparing our high precision numerical result with the asymptotic expansion
of the monopole mass for $\beta\ll1$ presented in Ref.\ \cite{Gardner82}
$$
M(\beta)=M_0\l( 1+\frac{1}{2}\,\beta
+\frac{1}{2}\,\beta^2\,\ln\beta+c_3\,\beta^2+\dots\r)\,,\qquad c_3=0.7071
$$
we have identified a discrepancy for the coefficient $c_3$, in that we have found
$c_3\approx0.4429$ from a numerical fit.
This discrepancy has led us to reexamine the computation and we
have found a computational error in \cite{Gardner82}.
Moreover we have found the following simple
analytic expression for the coefficient $c_3$:
$$
c_3=\frac{\ln(3\pi)}{2}-\frac{13}{24}-\frac{\pi^2}{72}=0.442926581...\,.
$$
We have also determined the first three coefficients of the asymptotic expansion
of the monopole mass in the limit $\beta \to \infty$, confirming and even improving on the
previous numerical results of \cite{Kirk-Zachos81}.

We believe that the numerical techniques developed in this paper will prove
useful to obtain good numerical precision for a much larger class of
nonlinear equations, where simpler methods are not readily applicable.

\section{The Monopole Equations}

The action of an $SU(2)$ Yang-Mills-Higgs theory with the Higgs field in the adjoint
representation is given as
\be\label{ActionYMH}
S_{\scriptstyle\rm YMH} = \int d^4x\;
\left(-\frac{1}{4\,e^2} F_{\mu\nu}^{a}F^{a\;\mu\nu}
+\frac{1}{2}(D_\mu\Phi^a)(D^\mu\Phi^a)-
\frac{\gamma}{8}{(\Phi^a \Phi^a-v^2)}^2\right)\;,
\ee
where $a=(1,2,3)$, $e$ is the gauge coupling constant,
$\gamma$ resp.\ $v$ denotes the coupling of the scalar self-interaction,
resp.\ the vacuum expectation value of the Higgs field.
The gauge field strength and the covariant derivative are defined as
\be
F_{\mu\nu}^{a} =\dr_{\mu}A_{\nu}^a - \dr_{\nu}A_{\mu}^a
+\epsilon^{abc}A_{\mu}^bA_{\nu}^c\,,\quad
D_{\mu}\Phi^a = \dr_{\mu}\Phi^a +\epsilon^{abc}A_{\mu}^b\Phi^c\;.
\ee
The mass of the Higgs, resp.\ of the gauge field
is given by $M_H=v\sqrt{\gamma}$, resp.\ $M_W=v\,e$.
The `minimal' static, purely magnetic, spherically symmetric Ansatz
for the gauge and Higgs fields can be written as
\be\label{AnsatzYMH}
A_{\scriptstyle 0}^a(r) = 0\,,\;\;
A_{\scriptstyle i}^a (r) = \epsilon_{iak}\frac{x_k}{r^2}\l(W(r)-1\r)\,,\;\;
\Phi^a(r) = H(r)\frac{x^a}{r}\,,
\ee
where $x^a$ are the cartesian coordinates and $r^2=x^k x_k$.
This Ansatz leads to the following reduced action:
\be\label{redactionYMH}
S_{\scriptstyle\rm YMH} = -\frac{4\pi v}{e}\int_0^{\infty} dr\;
\left( W'^2+\frac{r^2}{2}H'^2+
\frac{(1-W^2)^2}{2r^2}+\frac{\beta^2r^2}{8}(H^2-1)^2+W^2H^2\right)\,,
\ee
where the coordinate $r$ and the function $H$ have been
rescaled as
\be
r \rightarrow \frac{r}{ev}\,,\;\; H \rightarrow vH\,,\quad{\rm and}\quad\beta=M_H/M_W\,.
\ee
The equations of motion are obtained by varying the action
(\ref{redactionYMH}), and they can be recast as a non-autonomous dynamical system of the form
\bsub\label{YMHeqs}
W' & = U\,\\
U' & = & WH^2+\frac{W(W^2-1)}{r^2}\,,\\
H' & = & V\,,\\
V' & = & \frac{2HW^2}{r^2}+\frac{\beta^2}{2}H(H^2-1)-\frac{2 V}{r}\,,\
\esub
where $W$, $U$, $H$ and $V$ are now considered as independent phase-space variables.

Since the configuration is static, the self-energy, i.e.\
the mass of the monopole
is given by
$E=-S_{\scriptstyle YMH}$, and
to simplify matters we shall use the rescaled mass, $\tilde E$,
$$E=M_0\tilde E(\beta)\,,\quad {\rm where}\quad M_0=\frac{4\pi v}{e}\,.$$

Since we shall study in some detail the $\beta$ dependence of the mass function of
the monopole, it will prove useful
to compute the derivative of $\tilde E(\beta)$ w.r.\ to $\beta$, and one finds:
\be\label{dEdbeta}
\frac{d\tilde E(\beta)}{d\beta}=\frac{\partial\tilde E(\beta)}{\partial\beta}
= \frac{\beta}{4}\int_0^{\infty}\!dr\,r^2(H^2-1)^2 > 0\,,
\ee
where the equations of motion (\ref{YMHeqs}) have been used.
Equation (\ref{dEdbeta}) implies that the mass of the monopole increases monotonically with
increasing $\beta$.

Let us recall that for the case of a {\em massless} Higgs field, i.e.\
$\beta=0$, the solutions of Eqs.\ (\ref{YMHeqs}) are known in closed form
\cite{Bogomolnyi76,Prasad-Somm75}
 \be\label{PSsolution}
W_0(r)=\frac{r}{\sinh(r)}\,,\quad H_0(r)=\coth(r)-\frac{1}{r}\,,
\ee
which is the celebrated Bogomolnyi-Prasad-Sommerfield
(BPS) monopole solution.
In fact the BPS solution satisfies the following equations:
\be\label{Bogomeqs}
W_0'=-W_0H_0\,,\quad H_0'=\frac{1-W_0^2}{r^2}\,,
\ee
the spherically symmetric Bogomolnyi equations \cite{Bogomolnyi76}, which are much
simpler than the original field equations (\ref{YMHeqs}).
The (rescaled) mass of the BPS monopole is just $\tilde E(0)=1$.

In the limit $\beta \to \infty$ the potential energy term in (\ref{ActionYMH}) forces
the Higgs field to be frozen at its vacuum value almost everywhere ($H(r)\equiv 1$ $\forall r>0$, but
$H(0)=0$) and
 the equations of motion (\ref{YMHeqs}) reduce to a {\em massive} Yang-Mills equation
for the limiting solution $W_{\scriptscriptstyle\infty}(r)$
\be\label{EqDiffWbetainfty}
W_{\scriptscriptstyle\infty}''=W_{\scriptscriptstyle\infty}+
W_{\scriptscriptstyle\infty}\frac{W_{\scriptscriptstyle\infty}^2-1}{r^2}\,.
\ee
The mass of the monopole stays finite in this limit, and it is given by
\be\label{Einfinty}
\tilde E(\infty) = \int_0^{\infty} dr
\left[ W_{\scriptscriptstyle\infty}'^2+
\frac{(1-W_{\scriptscriptstyle\infty}^2)^2}{2r^2}+
W_{\scriptscriptstyle\infty}^2\right]\,.
\ee

\section{Numerical Techniques}

We briefly recall the local behaviour of the functions $H$ and $W$ for
$r\to0$ resp.\ $r\to\infty$, corresponding to
the two explicit singular points of the dynamical system (\ref{YMHeqs}a-d).
The boundary conditions at these points follow from
the requirement of regularity of the solutions,
which ensures the finiteness of the energy.

One can show that there exists a two-parameter family of
{\sl local} solutions regular at $r=0$ given by the expansion:
\be\label{bco}
W(r)=1-b\,r^2+O(r^4)\;,\;\;H(r)=a\,r+O(r^3)\,,
\ee
(as long as $\beta$ is finite) with free parameters $a$, $b$.
In fact from the finiteness of the energy (\ref{redactionYMH})
alone one can already deduce that
\be\label{bcinf}
H(r) \to1\,,\quad W(r) \to 0\,, \quad{\rm for}\quad r \to +\infty\,.
\ee
In order to obtain globally regular, finite energy solutions
of Eqs.\ (\ref{YMHeqs}a-d) by numerical integration
from the origin, one has to suppress the exponentially growing
modes present in Eqs.\ (\ref{YMHeqs}a-d) by
fine-tuning the parameters $a$,$b$ defined at $r=0$.
To exhibit the exponentially growing modes
in the equations of motion for $\beta\ne0$,
we introduce the following new variables
\be
W_+=W+U\;,\;W_-=W-U\;,\;h_+=\beta h+v\;,\;h_-=\beta h-v\,,
\ee
with $h=r(H-1)$ and $v=h'$, which satisfy the following dynamical system
\be\label{diageqdf}
\begin{array}{ll}
W_+'  =  W_+ + F_W(W,h,r)\;, & h'_+  = \beta\,h_+ + F_h(W,h,r)\,,\\
& \\
W_-'  =  -W_- - F_W(W,h,r)\;, & h'_-  = -\beta\,h_- - F_h(W,h,r)\,,\
\end{array}
\ee
where
\bsub\label{FWFh}
F_W(W,h,r) & = &
\frac{Wh}{r}(\frac{h}{r}+2)+\frac{W(W^2-1)}{r^2}\,,\\
F_h(W,h,r) & = &
\frac{2W^2}{r}(1+\frac{h}{r})+\frac{\beta^2h^2}{2r}(3+\frac{h}{r})\,.\
\esub
From Eqs.\ (\ref{diageqdf}) it is clear that $h_+$, $W_+$
correspond to exponentially growing modes as $r$ increases,
while $h_-$, $W_-$ are decreasing modes. Finiteness of the energy requires
the absence of the growing modes for $r\to\infty$, i.e.\  the functions,
$h_+$, $W_+$, should tend to zero for $r\to\infty$.
This implies that the asymptotic behaviour of $H$, $W$ regular for $r\to\infty$ is
given as
\bsub\label{Whatinf}
W(r) &=& C_{\scr W}\,e^{-r}\l(1+O(\frac{1}{r})\r)\,,\\
H(r) &=& 1-C_{\scr H}\,\frac{e^{-\beta r}}{r}\l(1+O(\frac{1}{r})\r)\,,
\quad{\rm for}\ \beta<2\,\\
H(r) &=& 1-\frac{2C_{\scr W}^2}{\beta^2-4}\frac{e^{-2r}}{r^2}\l(1+O(\frac{1}{r})\r)
+C_{\scr H}\,\frac{e^{-\beta r}}{r}\l(1+O(\frac{1}{r})\r)\,,
\quad{\rm for}\ \beta>2\,,
\esub
where $C_{\scr H}$, $C_{\scr W}$ are free parameters determined only by
global requirements like regularity at $r=0$. Note that for $\beta>2$ the asymptotic
behaviour of $H$ is dominated by that of $W$ induced by the nonlinear terms (compare
Eq.\ (\ref{Whatinf}c) with Eq.\ (27) of Ref.\ \cite{Kirk-Zachos81}).
For large values of $\beta$ the two divergent modes, i.e.\ that of the Higgs field
$\sim e^{\beta r}$, and that of $W$ ($\sim e^{r}$), are quite different
that is we have a stiff system.

Let us now sketch the numerical methods we have adapted.
Since the locally regular solutions at the singular point $r=0$
admit a power series expansion (\ref{bco})
with two undetermined parameters $a$ and $b$,
it is perfectly adequate to integrate the equations
(\ref{YMHeqs}a-d) from the origin
using standard Runge-Kutta methods.
Actually, we have implemented an adaptive stepsize fifth-order Runge-Kutta
algorithm \cite{NumRec}.

At $r=+\infty$, the locally regular solutions
do not admit, however, a power series expansion, therefore
we have chosen to use integral equations, which are well suited to
deal with such singular points.
The solution of the system (\ref{diageqdf}), satisfying the condition (\ref{bcinf})
is given by the following integrals
\bsub\label{inteq}
W_+(r)  & = &  \disp\int_{\infty}^r e^{r-r'}\,F_W(W,h,r')\,dr'\;, \\
W_-(r)  & = & W_-(r_m)e^{r_m-r}-\disp\int_{r_m}^r e^{r'-r}\,F_W(W,h,r')\,dr'\;,\\
h_+(r)  & = & \disp\int_{\infty}^r e^{\beta(r-r')}\,F_h(W,h,r')\,dr'\;, \\
h_-(r)  & = & h_-(r_m)e^{\beta(r_m-r)}-\disp\int_{r_m}^re^{\beta(r'-r)}\,F_h(W,h,r')\,dr'\,,\
\esub
for some $r_m > 0$.
In order to solve these integral equations numerically
we have compactified the `outer' region,
$[r_{m},+\infty)$, by introducing the variable $t=1/r$.
Then, the interval $[0,t_m=1/r_m]$ is discretized and the solution
is obtained by iteration using
\be
\begin{array}{ll}
W_+(r)  =  0, &
W_-(r)  =  W_-(r_m)e^{r_m-r}\,,\\
& \\
h_+(r)  = 0, &
h_-(r)  = h_-(r_m)e^{\beta(r_m-r)}\,.\
\end{array}
\ee
as an initial configuration.
Our general strategy to find globally regular solutions
is to fine-tune $a$ and $b$ together with
$W_-(r_m)$ and $h_-(r_m)$,
in order to match the functions $W$, $U$, $H$ and $V$ at $r=r_m$.

The above procedure works well and yields accurate results for
$10^{-2}\lesssim \beta\lesssim 7$. For values of $\beta<10^{-2}$ one
has to increase $r_m$ more and more to ensure convergence of the
iteration of the integral equations (\ref{inteq}a-d). This leads
to a loss of accuracy, however, mainly due to the increasing
length of the integration interval. To overcome this problem we
have implemented a suitable version of the `multi-shooting'
method, by introducing some intermediate domains between the inner
region (containing the origin) and the external one
($[0,t_m=1/r_m]$). We have integrated in each domain independently
and then performed a {\em global} matching to obtain a regular
solution by fine-tuning the set of $4\times(m+1)$ free parameters,
$\{a\,,b\,,W_-(r_m)\,,h_-(r_m)\,,
W(r_i)\,,U(r_i)\,,H(r_i)\,,V(r_i)\,,\vert\,i=0,\ldots,m-1\}$, where
$m$ denotes the number of domains. For small values of $\beta$,
the dominant divergent mode comes from the gauge sector $\sim
e^{r}$, and the choice of intermediate regions of roughly equal
size $\simeq 7$ has turned out to be adequate.

For large values of $\beta$, the situation changes radically,
since the dominant divergent mode then comes from
the Higgs sector and grows as $\sim e^{\beta\,r}$.
Runge-Kutta algorithms fail to cope with such
exponentially growing modes on intervals of size $\gg 1/\beta$.
Therefore one should introduce $m\sim\beta$ domains for large
values of $\beta$, leading to a huge increase
in CPU time and to potential numerical problems such as instabilities.
To overcome this problem
we have implemented a simple version of multi-domain spectral methods
which are of {\em global} nature and are known for their high accuracy,
see Ref.\ \cite{lorene} for their use in Numerical Relativity.
These techniques have turned out to be
quite suitable to deal with such exponentially growing modes.
In the procedure we have implemented,
the values of $W_+$, $W_-$, $h_+$ and $h_-$ are iteratively computed
starting from an initial guess. In each domain we parametrize the
solutions by the values of $W_-$ and $h_-$ at $r_{i}$ and $W_+$ and $h_+$ at $r_{i+1}$
in order to control the growing modes during the iteration.
We describe briefly some of the basic ideas
of spectral methods in Appendix $1$.

The two cases, $\beta=0$ and $\beta=\infty$, require some special considerations.
When $\beta=0$
the asymptotic behaviour of the BPS solution
is dominated by the long range interaction of the massless Higgs field
and the generic behaviour of $H$ at infinity is
\be\label{PSatinf}
H(r) = d_0+\frac{d_1}{r}+o(r^{-2})\,.
\ee
where $d_0$, $d_1$ are free parameters.
The suitable combinations in the Higgs sector are then given by
\be
h_+=H-1+r\,V\;,\quad h_-=r^2\,V\,,
\ee
and the BPS solution (\ref{PSsolution}) is obtained by imposing the boundary conditions
$W_+=h_+=0$ at infinity.

In the limit $\beta\to\infty$, one has to integrate
the massive Yang-Mills equation (\ref{EqDiffWbetainfty}).
Due to the singular behaviour of the Higgs
field for $\beta=\infty$, the function $W_{\scr{\infty}}$ remains differentiable but not $C^2$
at the origin
\be\label{Winforigin}
W_{\scriptscriptstyle\infty}(r)= 1+\frac{1}{3}r^2\ln(r)-b'_{\infty}\,r^2+O(r^4\ln(r))\,.
\ee
The behaviour of $W_{\scr{\infty}}$ at infinity is not modified and one has to solve
the integral equations (\ref{inteq}a-b) with $F_W(W_{\scr{\infty}},h\equiv 0,r)$.

\section{Numerical Results}

\begin{figure}[ht]
\hbox to\linewidth{\hss%
  \psfrag{X}{$\log(1+r)$}
    \resizebox{8cm}{6.5cm}{\includegraphics{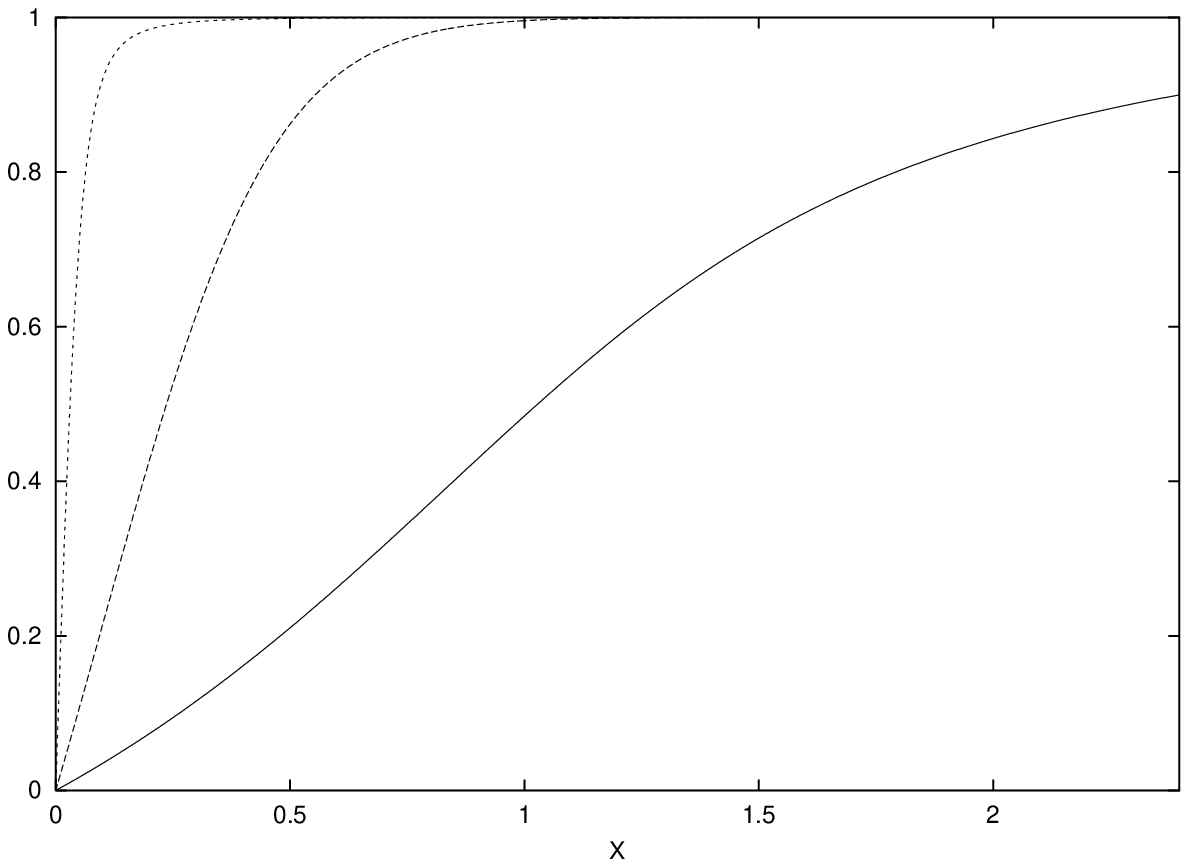}}%
\hspace{5mm}%
  \psfrag{r}{$r$}
    \resizebox{8cm}{6.5cm}{\includegraphics{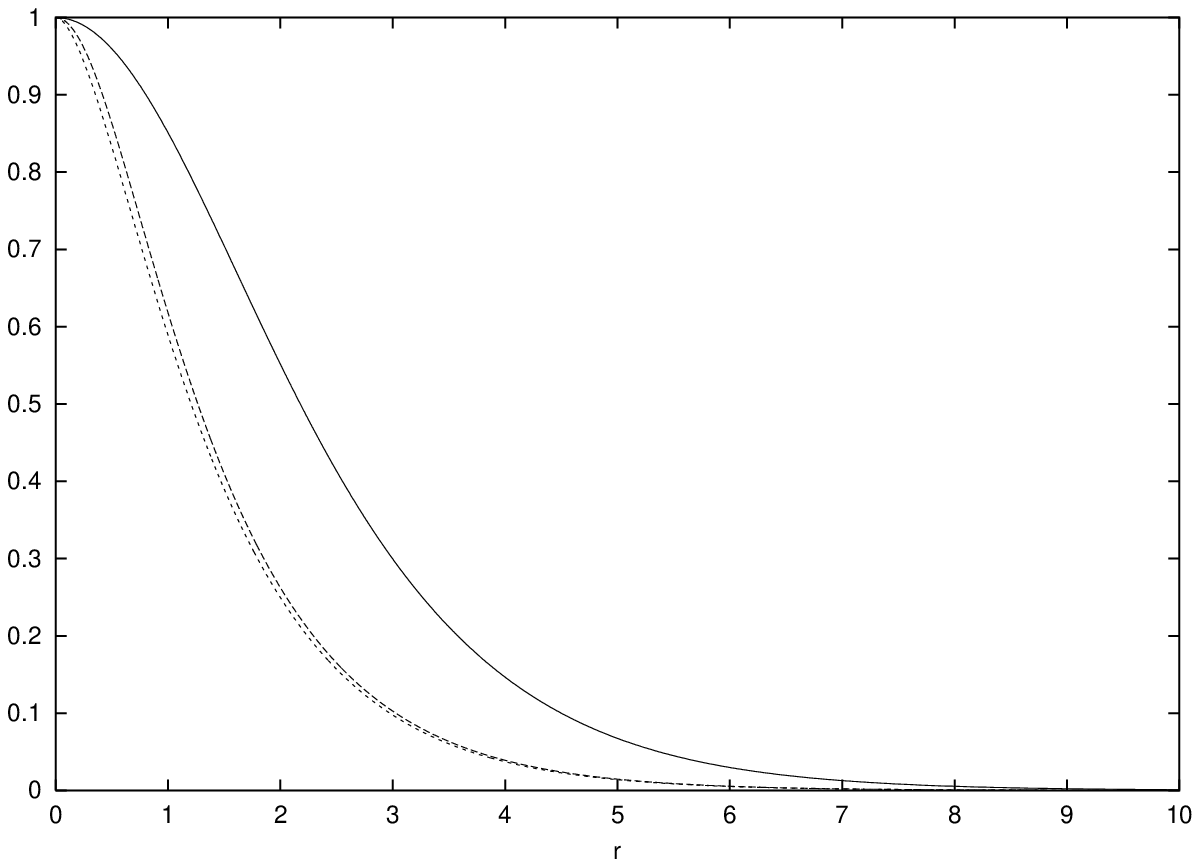}}%
\hss}
\caption{Plots of $H$ and $W$ for $\beta=0$ (solid), $5$ (dashed) and $50$ (dotted).
Note the different scales for $H$ and for $W$.}
\label{plotWH}
\end{figure}

We present the numerically obtained values
for the parameters $a$ and $b$ as well
as for the energy, $\tilde E$, in Tables $1$, $2$ and $3$. Figures \ref{plotWH} show
$H$ and $W$, while Figure $2$
displays the mass function, $\tilde E(\beta)$ for $0\le\beta\le 50$. We remark that our
numerical data give strong evidence that $H$ resp.\ $W$ are
monotonically increasing resp.\ decreasing functions of the radial variable, $r$,
for all values of $\beta$.

We expect that our numerical results for $a$, $\tilde E$, are accurate up to
11 significant digits. The parameter $b$ is somewhat less accurate for large
values of $\beta$: its accuracy is around 9 significant digits.
The expectation for the quoted accuracy is based on the stability of the results
w.r.\ to any change of the control parameters of the numerical integration.
We shall use our high precision numerical results to extract the
asymptotic behaviour of the monopole mass, $\tilde E(\beta)$ for $\beta\to0$ and
for $\beta\to\infty$.
\begin{figure}[ht]
\hbox to\linewidth{\hss
\psfrag{beta}{$\beta$}
\resizebox{12cm}{6.5cm}{\includegraphics{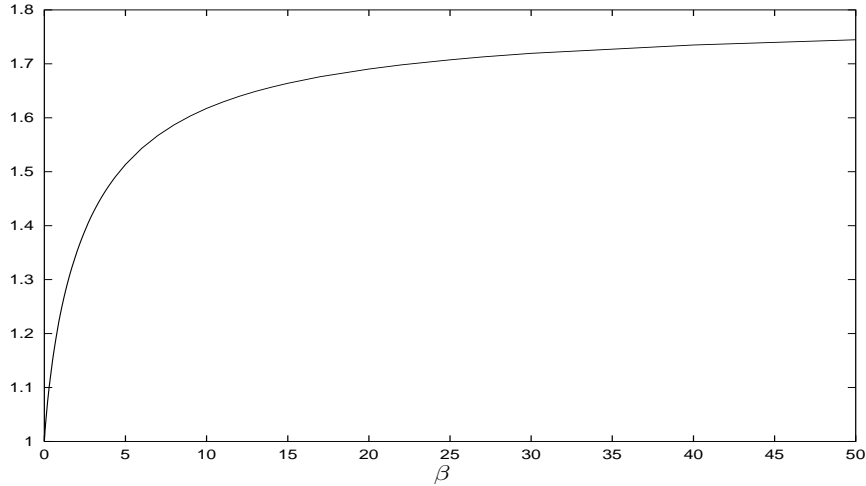}}
\hss}
\caption{The monopole mass, $\tilde E(\beta)$, plotted against $\beta$. }
\label{plotE}
\end{figure}

We study first the behaviour of the monopole mass for values of $\beta\ll1$.
The analytically known value of the energy for $\beta=0$,
$\tilde E(0)=1$, provides a good test for the numerical results.
In fact we have obtained numerically $\tilde E(0)=1.0$ to an
accuracy better than $10^{-11}$.
As it has been shown in Ref.\ \cite{Gardner82} for $\beta\ll1$ the monopole mass,
$\tilde E(\beta)$, can be expanded as
\be\label{Eexpansion}
\tilde E(\beta) = c_0+c_1\,\beta
+c_2\,\beta^2\,\ln\beta+c_3\,\beta^2+o(\beta^2\,\ln\beta)\,.
\ee
To determine the coefficients $c_i$ from our results, we have
computed $\tilde E(\beta)$
for $81$ values of $\beta$ in the range $10^{-4}\le\beta\le 5\times10^{-4}$, and
performed a standard least square fit, implementing a $\chi^2$-fitting routine
as described in Ref.\ \cite{NumRec}.
We have first minimized $\chi^2$ to obtain an
estimate for the values of the coefficients $c_i$, and then
determined the corresponding mean standard deviations, $\sigma_{\tilde E}$
and $\sigma_{c_i}$ of $\tilde E(\beta)$
respectively of the estimated values of the coefficients $c_i$.
This way we have obtained
\beqn\label{fitE}
c_0 &=& 1 \pm 8\times10^{-11}\,,\qquad c_1= 0.4999992 \pm 1.2\times10^{-7}\,,
\nonumber\\
c_2&=&0.496 \pm 2\times10^{-3}\,,\quad
c_3=0.41 \pm 1.5\times10^{-2}\,,\quad\sigma_{\tilde E}\simeq 3\times10^{-11}\,.
\eeqn
The values of the coefficients $c_0$, $c_1$, $c_2$ are all in
good agreement with the theoretical results of \cite{Gardner82,Kirk-Zachos81}.
The value, $c_3=0.41$, however, deviates quite significantly from the
one found in Ref.\ \cite{Gardner82}, where
the value of $c_3=0.7071$ has been quoted.
In order to improve upon the precision of the value of $c_3$ we have also made a
fit on
\be\label{Eexpansion-num}
\tilde E(\beta) = 1+\frac{1}{2}\,\beta
+\frac{1}{2}\,\beta^2\,\ln\beta+c_3\,\beta^2+d\,\beta^3\,\ln\beta\,,
\ee
i.e.\ fixing $c_0$, $c_1$, $c_2$ at their theoretical values, and
moreover we added an extra term, $d\,\beta^3\ln\beta$,
to parametrize the contribution of the unknown higher order corrections.
This way we have obtained
\be\label{fitc3}
c_3=0.44297\pm 1.8\times10^{-4}\,,\;\;d=0.64\pm 5\times10^{-2}\,,\;\;
\sigma_{\tilde E}\simeq 3\times10^{-13}\,,
\ee
a value fully consistent (within $2\sigma_{c_3}$) with the one found in Eq.\ (\ref{fitE}).
Without the ``improvement term'', $d\,\beta^3\ln\beta$, the coefficient in question is given as
$c_3=0.440890\pm 3.4\times10^{-5}$.
Being confident as to the correctness of the numerical results,
we have been led to reconsider the analysis of Ref.\ \cite{Gardner82};
this will be outlined in the next section.

Next we consider the asymptotic behaviour of $\tilde E$ for $\beta\gg 1$.
Motivated by the results of Ref.\ \cite{Kirk-Zachos81} we postulate
the following asymptotic expansion for $\tilde E$:
\be\label{Eexpatinf}
\tilde E = e_{\scr0} + \frac{e_{\scr-1}}{\beta} + \frac{e_{\scr-2}}{\beta^2}+
\frac{e_{\scr-3}}{\beta^3} +o(\beta^{-3})\,.
\ee
Following the same procedure as for $\beta\ll1$, we have determined the
numerical values of the coefficients $e_{-i}$ from a fit on $21$ values
in the range $1000\leq\beta\leq 2000$ :
\beqn\label{fitEinf}
e_{\scr0}&=&1.786658424\pm 1.7\times10^{-10}\,,\quad
e_{\scr-1}=-2.228956\pm 7.2\times10^{-6}\,,\nonumber\\
e_{\scr-2}&=&7.14\pm 1\times10^{-2}
\,,\quad e_{\scr-3}=-32.7\pm 4.5\quad \sigma_{\tilde E}\simeq 3\times10^{-11}\,.
\eeqn
These results provide some motivation
to prove the validity of the expansion (\ref{Eexpatinf}) by analytic methods.
The value of $\tilde E(\infty)$ (corresponding to $e_{\scr0}$) has been obtained
{\em independently}
from Eq.\ (\ref{Einfinty}) and is given in
Table $3$. Note the good agreement between these two values.

Some comments are in order on the obtained precision for the values of
$a$, $b$, and on their respective $\beta$-dependence.
Since for $\beta\gg1$ the Higgs field rises steeply
in an interior region of size $\sim1/\beta$ close to its vacuum expectation
value (see also Fig.\ \ref{plotWH}),
$H$ approaches more and more a step-function as $\beta$ increases.
Hence, in order to numerically integrate the equations near the origin,
one has to choose an interior region of size $\sim 1/\beta$.
Such a choice enables us to determine the parameter $a$ with a good accuracy.
Indeed, $H$ is a linear function of $r$ within some interval $[0,C/\beta]$,
therefore we expect the dominant term of $a$ to be a linear function of $\beta$.
Indeed, taking
\be\label{a_beta}
a(\beta) \simeq a_1\,\beta +a_0+\frac{a_{\scr-1}}{\beta}+\frac{a_{\scr-2}}{\beta^2}\,,
\ee
and fitting the coefficients we have found:
\beqn\label{a_beta_coeff}
a_1&=&0.357826169\pm 2\times10^{-9}\,,\quad a_{\scr0}=0.000677\pm 7.5\times10^{-6}\,,
\nonumber\\
a_{\scr-1}&=&6.37\pm 10^{-2}\,\quad a_{\scr-2}=-440\pm 5\,,\quad
\sigma_{a}\simeq 8.2\times10^{-9}\,,
\eeqn
for values of $\beta$ ranging from $1000$ to $2000$.

Related to the above choice for the interior region there is, however,
some loss of precision on the value of $b$.
This is due to the fact that $W$, as opposed to $H$, changes very little
near the origin. We have found that the precision we get for $b$ is only
$10^{-9}$. This loss in precision for the parameter $b$ does not
significantly affect the precision obtained for the energy, which is
an integrated quantity.
Concerning the behaviour of $b$ for large $\beta$, we would like
to point out that $b'_{\infty}$ as defined in (\ref{Winforigin})
does not correspond to the limit of $b$ as $\beta\rightarrow\infty$.
That is why the value $b'_{\infty} = 0.4843164140$
 is absent from Table $3$.
Indeed for $\beta\to\infty$ the parameter $b$ diverges (similarly to $a$),
leading to the $r^2\,\ln(r)/3$ term in (\ref{Winforigin}). However,
we expect its divergence to be weaker than that of $a$. We found
the behaviour of $b$ to be well approximated by
\be\label{b_beta}
b(\beta) \simeq (0.33333270\pm 2\times10^{-8})\,\ln\beta+
(0.065700\pm 2\times10^{-6})
+(6.82\pm1.8\times10^{-3})\,\beta^{-2}\,,
\ee
with $\sigma_{b}\simeq3\times10^{-9}$.
We note that the numerical value of the coefficient of the $\ln\beta$ in
(\ref{b_beta}) is quite close to $1/3$
which is precisely the coefficient of the $r^2\,\ln(r)$ term in (\ref{Winforigin}).
It is very natural to conjecture that $b(\beta)\sim\,(1/3)\ln\beta$ as
$\beta\to\infty$.
\begin{table}[ht]
\parbox{12cm}{
\caption{Numerical values of $a$, $b$ and $\tilde E$ for $0\leq \beta\leq 10^{-3}$.
$N$ denotes the number of domains of integration including the interior region
and the exterior compactified region containing $r=\infty$.}}
 \vskip 0.5cm
\begin{tabular}{|c|c|c|c|c|} \hline
$\beta$ & $a$ & $b$ & $\tilde E$ & $N$\\
\hline
0      & 0.3333333333 & 0.1666666667 & 1.0000000000 & 2\\
\hline
0.0001 & 0.3333999248 & 0.1666999621 & 1.0000499584 & 5\\
0.0002 & 0.3334663934 & 0.1667331954 & 1.0000998473 & 5\\
0.0003 & 0.3335327549 & 0.1667663745 & 1.0001496747 & 5\\
0.0004 & 0.3335990174 & 0.1667995035 & 1.0001994446 & 5\\
0.0005 & 0.3336651866 & 0.1668325852 & 1.0002491600 & 3\\
0.0006 & 0.3337312672 & 0.1668656218 & 1.0002988231 & 3\\
0.0007 & 0.3337972625 & 0.1668986152 & 1.0003484357 & 3\\
0.0008 & 0.3338631756 & 0.1669315669 & 1.0003979993 & 3\\
0.0009 & 0.3339290091 & 0.1669644781 & 1.0004475152 & 3\\
0.0010 & 0.3339947653 & 0.1669973500 & 1.0004969846 & 3\\
\hline
\end{tabular}
\end{table}
\begin{table}[ht]
\parbox{12cm}{
\caption{Numerical values of $a$, $b$ and $\tilde E$ for $10^{-2}\leq \beta\leq 7$.
For this range of values of $\beta$, only $2$ domains are needed. }}
 \vskip 0.5cm
\begin{tabular}{|c|c|c|c|} \hline
$\beta$ & $a$ & $b$ & $\tilde E$\\
\hline
0.01 & 0.3396968769 & 0.1698451903 & 1.0048108392\\
0.05 & 0.3624466559 & 0.1811436175 & 1.0220791858\\
0.1  & 0.3878538429 & 0.1936144524 & 1.0410559503\\
0.5  & 0.5533233202 & 0.2697154617 & 1.1493759746\\
1    & 0.7318137699 & 0.3409021837 & 1.2377010395\\
2    & 1.0683224404 & 0.4490865847 & 1.3510046819\\
3    & 1.4003124276 & 0.5320943027 & 1.4236384997\\
4    & 1.7338856977 & 0.5997114280 & 1.4749928035\\
5    & 2.0701446040 & 0.6567291699 & 1.5134382751\\
6    & 2.4090593765 & 0.7059674601 & 1.5433692766\\
7    & 2.7503251769 & 0.7492501033 & 1.5673591187\\
\hline
\end{tabular}
\end{table}
\begin{table}[ht]
\parbox{8.5cm}{
\caption{Numerical values of $a$, $b$ and $\tilde E$ for $\beta\ge 10$.}}
 \vskip 0.5cm
\begin{tabular}{|c|c|c|c|c|} \hline
$\beta$ & $a$ & $b$ & $\tilde E$ & $N$\\
\hline
10    & 3.7849823090 & 0.8542424691 & 1.6173762372 & 3\\
50    & 17.964403588 & 1.371251333  & 1.7447435157 & 4\\
100   & 35.825856171 & 1.60118232   & 1.7650569743 & 4\\
500   & 178.92488493 & 2.13720411   & 1.7822288542 & 4\\
1000  & 357.83277188 & 2.36823357   & 1.7844365846 & 5\\
1100  & 393.61488618 & 2.40000240   & 1.7846379843 & 7\\
1200  & 429.39707896 & 2.42900524   & 1.7848059066 & 7\\
1300  & 465.17933308 & 2.45568539   & 1.7849480580 & 7\\
1400  & 500.96163610 & 2.48038744   & 1.7850699484 & 7\\
1500  & 536.74397876 & 2.50338457   & 1.7851756213 & 7\\
1600  & 572.52635402 & 2.52489700   & 1.7852681115 & 7\\
1700  & 608.30875644 & 2.54510487   & 1.7853497407 & 7\\
1800  & 644.09118172 & 2.56415738   & 1.7854223163 & 7\\
1900  & 679.87362645 & 2.58217954   & 1.7854872652 & 7\\
2000  & 715.65608785 & 2.59927709   & 1.7855457295 & 7\\
\hline
$\infty$&    -       &     -        & 1.7866584230 & 2\\
\hline
\end{tabular}
\end{table}
\section{Asymptotic analysis of the monopole mass}

In this section we study the mass of the monopole in function
of the mass ratio $\beta=M_{\rm H}/M_{\rm W}$ for the two extreme cases
$\beta\ll1$ (i.e.\ near the BPS limit) and for $\beta\gg1$ (i.e.\ near the massive
Yang-Mills limit).

We present first a calculation of the monopole energy for $\beta\ll1$ to order
$\beta^2$.
This computation has been first performed in \cite{Gardner82}
based on matched asymptotic expansion techniques.
The discrepancy between our numerical results presented in Section 4
and those predicted in \cite{Gardner82} has led us to repeat the computation
of \cite{Gardner82}.

It is natural to expand the monopole solution for $\beta\ll1$ around the BPS solution
as $H_{\rm N}=H_0+\delta H\,,W_{\rm N}=W_0+\delta W$ and try to determine
$\delta H\,,\delta W$,
by linearizing the field equations (\ref{YMHeqs}a-d).
The small perturbations  $\delta H$, $\delta W$
should satisfy {\em regular} boundary conditions at the {\em origin}.
%and must have {\em exponential decay} for $r\to\infty$
This approximation is valid, however, only in an interval $0\le r \le \beta^{-\alpha_1}$ with some
$\alpha_1<1$ (`near' region), but breaks down in the `far' region,
$r\gg\beta^{-\alpha_1}$, where the Higgs field decays exponentially,
$H=H_{\rm F}\approx 1+Ae^{-\beta r}/r+\ldots$
with some constant $A$, as opposed to the slow $1/r$ falloff of the massless BPS solution.
There is, however, an `intermediate' region,
$\beta^{-\alpha_2}\le r \le\beta^{-\alpha_1}$, $0<\alpha_2<\alpha_1<1$
where both approximations hold, and one can match $H_{\rm N}$
with $H_{\rm F}$.
(The actual values of the exponents, $\alpha_1$, $\alpha_2$, do not (and should not) matter
for the computation. In \cite{Gardner82} the values $\alpha_1=1/2$, $\alpha_2=1/4$
have been chosen for concreteness.)
This matching is possible
using the remaining free parameters in $ H_{\rm N}\,,W_{\rm N}$ resp.\ $H_{\rm F}\,,W_{\rm F}$
after having imposed regularity at $r=0$, resp.\ at $r=\infty$.

We rewrite the field equations (\ref{YMHeqs}a-d) introducing the rescaled variable
$x=\beta\,r$, which will prove useful in the far region and also for the $\beta\gg1$ limit:
\bsub\label{YMHeqsx}
\ddot{W} & = & W\left(\frac{H}{\beta}\right)^2+\frac{W(W^2-1)}{x^2}\,,\\
\ddot{H} & = & 2H\left(\frac{W^2}{x^2}+\frac{1}{4}(H^2-1)\right)-\frac{2\dot{H}}{x}\,,\
\esub
where $\dot{f}=df/dx$.

In the far region, $x\gg1$,
the field equations (\ref{YMHeqsx}) decouple since both fields fall off exponentially.
$H_{\rm F}(x)$ then satisfies (\ref{YMHeqsx}b) with $W\equiv0$.
Expanding $H_{\rm F}(x)$ around the free massive solution as
\be\label{Hfar}
H_{\rm F} = 1+A_{\rm F}\,\frac{e^{-x}}{x}+A_{\rm F}^2\,\frac{f_1(x)}{x}\,e^{-2\,x}+A_{\rm F}^3\,\frac{f_2(x)}{x}\,e^{-3\,x}
+\ldots\,,
\ee
where $A_{\rm F}$ is the (small) expansion parameter, one obtains a set of
linear differential equations for $f_n(x)$ which can be solved recursively. To determine $c_3$
one only needs in fact $f_1(x)$, which satisfies:
 \be
 \ddot{f_1}-4\dot{f_1}+3f_1=\frac{3}{2x}\,,
 \ee
 together with the boundary condition $f_1(x)\to0$ as $x\to\infty$.
 The solution for $f_1(x)$ is then given as
\be
f_1(x) = \frac{3}{4}\left(e^{3\,x}\,\mbox{Ei}(-3\,x)-e^x\,\mbox{Ei}(-x)\right)\,, \quad
{\rm with}\quad \mbox{Ei}(-x) = \int_{\infty}^{x}\,dt\,\frac{e^{-t}}{t}\,.
\ee
Now it is not difficult to determine $H_{\rm F}(r)$ in the intermediate region,
where $x=\beta r\ll1$, containing $A_{\rm F}$ as a parameter to be determined
order by order in $\beta$,
\be\label{Hfarexp}
H_{\rm F}(r)=1+\frac{A_{\rm F}}{\beta r}\l(1+\frac{3\ln3}{4}A_{\rm F}\r)+\frac{A_{\rm F}\beta}{2}
\l(\frac{A_{\rm F}}{\beta}3\ln r+r\r)+B_{\rm F}+o(\beta^2)\,,
\ee
with
$$
 B_{\rm F}= A_{\rm F}\l[-1+\frac{3A_{\rm F}}{2}
\l(\frac{\ln3}{2}+\ln\beta+\gamma_{\rm E}-1\r)\r]\,.
$$
The above explicit form of $B_{\rm F}$ will not be needed
for the calculation of the monopole mass, only the fact that $B_{\rm F}$ does not depend on $r$.

In the near region it is somewhat more difficult to determine the solution, because
the field equations are coupled there. The Green function in the
BPS monopole background can be explicitly computed
(using the Bogomolnyi equations) see e.g.\ Ref.\ \cite{Rossireview}.
This remarkable fact has been well exploited in \cite{Gardner82}
to find the solution $H_{\rm N}$, $W_{\rm N}$.
We omit here the somewhat lengthy computational details, since we basically agree with the
results of Ref.\ \cite{Gardner82},
and give directly the final result for the expansion of $H_{\rm N}$ in the
intermediate region:
\be\label{Hnear}
H_{\rm N}(r)=1-\frac{1}{r}+\frac{\beta^2}{2}(3\ln r-r-3)+\frac{\beta^2}{2}\l(3\ln \Lambda+
\frac{2}{\Lambda}\r)+A_{\rm N}(\Lambda,\beta)+ o(\beta^2)\,,
\ee
where $r<{\Lambda}\leq\beta^{-\alpha_1}$ is a kind of `cutoff', which can be
arbitrarily chosen within the intermediate region,
 $A_{\rm N}(\Lambda,\beta)$ is the matching parameter.
We remark that in contradistinction to $A_{\rm F}$ in
the development of $H_{\rm F}$,  $A_{\rm N}(\Lambda,\beta)$ turns out to depend
{\em explicitly} on the chosen limit of the intermediate region, $\Lambda$.
The value of $A_{\rm F}$ is
now easily found by matching the $r$-dependent terms in $H_{\rm F}(r)$, Eq.\ (\ref{Hfarexp})
and in $H_{\rm N}(r)$, Eq.\ (\ref{Hnear}), yielding
$$
A_{\rm F}=-\beta-\frac{3}{4}(\ln 3)\,\beta^2+O(\beta^3)\,,\quad
B_{\rm F}=\beta\l[1+\frac{3\beta}{2}\l(3\ln\l(3\beta\r)+\gamma_{\rm E}-1\r)\r]
+ o(\beta^2)\,.
$$
The value of $A_{\rm N}(\Lambda,\beta)$ is then immediately found from the above results.
%$$
%A_{\rm N}= -\frac{\beta^2}{2}\l(3\ln \Lambda+\frac{2}{\Lambda}\r)+
%\beta\l[1+\frac{3\beta}{2}\l(3\ln\l(3\beta\r)+\gamma_{\rm E}\r)\r]
%+ o(\beta^2)\,.
%$$
Note the {\em nonanalytic dependence} of $H_{\rm F}(r)$ and $H_{\rm N}(r)$
on $\beta=\sqrt{\beta^2}$ and the appearance of terms $\propto\ln r$, ultimately responsible
for correction terms of the type $\beta\ln\beta$ to the monopole mass.
%We also remark that $A_{\rm F}$ is completely determined by comparing the terms proportional to
%$1/r$ in $H_{\rm F}(r)$ and $H_{\rm N}(r)$.

Now to compute the monopole mass for small values of $\beta$ we evaluate first
$d{\tilde E}/d\beta$ (\ref{dEdbeta})
from which one easily obtains $\tilde E(\beta)$ integrating w.r.\ to \ $\beta$.
By splitting the integral in Eq.\ (\ref{dEdbeta}) at some matching point $r_{\rm m}$,
such that $\beta^{-\alpha_2}<r_{\rm m}<\Lambda$,
\be
\frac{d{\tilde E}}{d\beta}= {\tilde E}^{'}_{\rm N}(\beta,r_{\rm m})+
{\tilde E}^{'}_{\rm F}(\beta,r_{\rm m})\,,
\ee
one can see that to first order in $\beta$ the two pieces
${\tilde E}^{'}_{\rm N}(\beta,r_{\rm m})$, resp.\
${\tilde E}^{'}_{\rm F}(\beta,r_{\rm m})$
are given as
\be\label{Energy-cut}
{\tilde E}^{'}_{\rm N}(\beta,r_{\rm m})=\frac{\beta}{4}\int_{0}^{r_{\rm m}}dr\,r^2(H_0^2-1)^2\,,
\qquad
{\tilde E}^{'}_{\rm F}(\beta,r_{\rm m})=
\frac{\beta}{4}\int_{r_{\rm m}}^{\infty}dr\,r^2(H_{\rm F}^2-1)^2\,,
\ee
with $H_{\rm F}$ taken to order $A_{\rm F}^2$ in Eq.\ (\ref{Energy-cut}).
For the details of the computation of
$d{\tilde E}/d\beta$ we refer to Appendix 2.
From the result (\ref{exp4}) found in Appendix 2 we obtain the development of the monopole mass
for small $\beta$:
\be
\tilde E(\beta) = 1+
\frac{1}{2}\beta+\frac{1}{2}\beta^2\ln\,\beta+
\frac{1}{2}\left(\ln(3\pi)-\frac{13}{12}-\frac{\pi^2}{36}\right)\beta^2+o(\beta^2)\,.
\ee

Next we consider the $\beta\rightarrow\infty$ limit, already analyzed in detail
in Ref.\ \cite{Kirk-Zachos81}. As shown in \cite{Kirk-Zachos81}
the mass function can be expanded as
\be
\tilde E(\beta)=\tilde E(\infty)+
\frac{1}{\beta}\frac{d\tilde E(\beta)}{d(1/\beta)}\Bigm|_{\beta=\infty}+O(1/\beta^2)\,.
\ee
The physical interpretation of this expansion is that the Higgs field does
not contribute to the energy in the limit $\beta\to\infty$.
Now, to evaluate $d{\tilde E}/d(1/\beta)$ at $\beta=\infty$
we take directly the $\beta\to\infty$ limit of Eqs.\ (\ref{YMHeqsx}) which
amounts to a `blow up' of the origin to an infinite interval corresponding to
the near region, where $W\equiv1$.
Therefore the equation for $H_{\infty}$ is simply
\be\label{Hcorreq}
\ddot{H}_{\infty}=2H_{\infty}\left(\frac{1}{x^2}+\frac{1}{4}(H_{\infty}^2-1)\right)-
\frac{2\dot{H}_{\infty}}{x}\,.
\ee
Note that $H_{\infty}$ corresponds to the interior of a global monopole.
In the  outside (`far') region (i.e.\ $r>0$), $H_{\infty}\equiv1$ and
$W_{\infty}$ satisfies the massive Yang-Mills equation (\ref{EqDiffWbetainfty}).

The first correction to the mass is of order $1/\beta$, and it is given by
\be\label{coeff_beta_inf}
\frac{d\tilde E(\beta)}{d(1/\beta)}\Bigm|_{\beta=\infty}
=-\int_0^{\infty}\,dx\,\frac{x^2}{4}(H_{\infty}^2-1)^2\,.
\ee
In order to compute this integral, one has to determine $H_{\infty}(x)$
by numerical integration of Eq.\ (\ref{Hcorreq}).
The regular boundary conditions at the origin resp.\ at infinity are
\be\label{Hbehaviour}
H(x) = a_{\infty}'\,x+O(x^3)\,,\quad{\rm resp.}\quad
H(x) = 1-\frac{2}{x^2}
-\frac{6}{x^4}-\frac{92}{x^6}+\dots+B\frac{e^{-x}}{x}\,,
\ee
%\mbox{and} \;\; H(x) &=& 1-\frac{2}{x^2}
%-\frac{6}{x^4}-\frac{92}{x^6}+\dots+B\frac{e^{-x}}{x}\,,
%\esub
%
where $a_{\infty}'$ and $B$ are a free parameters.
Note that the behaviour of $H(x)$ as $x\to\infty$
is dominated by an (asymptotic) series in $1/x$ with all coefficients determined,
the free parameter, $B$, is in the sub-dominant term, $e^{-x}/x$.
The numerical value of the coefficient (\ref{coeff_beta_inf}) is found to be
\be
\frac{d\tilde E(\beta)}{d(1/\beta)}\Bigm|_{\beta=\infty}=-2.2289514397\,,
\ee
which is in very good agreement with the value obtained from our fit
for $e_{-1}$. It also agrees with the result found
in Ref.\ \cite{Kirk-Zachos81}.
The value of $a_{\infty}'=0.3578262475$ agrees also quite well with
$a_1$ in Eq.\ (\ref{a_beta_coeff}). This is due to the fact that
$H \sim a_{\infty}'\,x=a_{\infty}'\,\beta\,r$,
therefore $a(\beta)\sim a_{\infty}'\,\beta$.
%Here, we want to draw the reader's attention to the fact that
%
%\be
%H \sim a_{\infty}'\,x=a_{\infty}'\,\beta\,r\,,
%\ee
%
%such that $a(\beta)\sim a_{\infty}'\,\beta$ as $\beta\to\infty$,
%in agreement with the previous result (\ref{a_beta}).

\section{Conclusion}

We have illustrated the usefulness of multi-domain spectral methods
combined with traditional Runge-Kutta integration
to obtain high precision numerical results for the venerable
't Hooft-Polyakov monopole.
As an application we have extracted the asymptotic behaviour of the monopole mass
for small and large values of $\beta$. Our numerical results for
small values of $\beta$ are in disagreement with
a previous result \cite{Gardner82}. Repeating the asymptotic analysis
of Ref.\ \cite{Gardner82} we have found a simple analytic expression
for the monopole mass, which is in full agreement with our numerical results.
For large values of $\beta$ our numerical results agree with those
of Ref.\ \cite{Kirk-Zachos81}.

%%%%%%%%%%%%%%%%%%%%%%%%%%%%%%%%%%%%%%%%%%%%%%%%%%%%%%%%%%%%%%%%%%%%%%%%%%%%%%%%

%%%%%%%%%%%%%%%%%%%%%%%%%%%%%%%%%%%%%%%%%%%%%%%%%%%%%%%%%%%%%%%%%%%%%%%%%%%%%%%%

\section*{Appendix 1}

We present here some basics of spectral methods used in our approach.
The fundamental idea underlying all spectral methods is to approximate a function $f$ as
\be
f(x) \simeq P^N(x) \equiv \sum_{j=0}^{N-1} \, c_j P_j(x)\,,
\ee
where the $P_j(x)$ form an orthonormal basis of polynomials of degree $\le (N-1)$
and where $f(x_k)=P^N(x_k)$ for a certain set of $N$ values $(x_0,...,x_{N-1})$,
called collocation points.
We have chosen Chebyshev polynomials, $T_j(x)$,
for the basis,  $P_j(x)$.
Since they are defined in the interval $[-1,1]$
by the relation $T_j(\cos(x))=\cos(j\,x)$,
we shall consider the spectral decomposition of functions defined in this
same interval.
Each Chebyshev polynomial, $T_j$, has $j$ zeros in the interval $[-1,1]$.
A natural choice of the $N$ collocation points where
the interpolating polynomial coincides with the values of the function $f$,
correspond to the $N$ zeros of $T_N(x)$ which are
\be
x_k=\cos\left(\frac{\pi(k+1/2)}{N}\right)\,,\;k=0,1...,N-1\,.
\ee
The Chebyshev polynomials satisfy the following discrete
orthogonality relation
\be
\sum_{k=0}^{N-1} T_i(x_k)T_j(x_k)=\left\{\begin{array}{l@{\quad \quad}l}
  0    &   i \neq j  \\
  N/2  &  i=j \neq 0 \\
  N    &  i=j=0
\end{array} \right.\ee
Defining the coefficients $c_j\,,\;j=0,1,\ldots,N-1$ by
\beqn\label{coeff}
c_j & = & \frac{2}{N}\sum_{k=0}^{N-1}\,f(x_k)\,T_j(x_k)\,,\nonumber\\
    & = & \frac{2}{N}\sum_{k=0}^{N-1}\,f(x_k)\,\cos\left(\frac{\pi j(k+1/2)}{N}\right)\,,\
\eeqn
the function $f$ matches its polynomial approximation
on the collocation points according to
\be\label{f}
f(x_k) = \frac{c_0}{2} + \sum_{j=1}^{N-1} \, c_j T_j(x_k)\,.
\ee
It is interesting to note that the expressions (\ref{coeff}) and (\ref{f})
can be computed numerically using FFT algorithm if $N$ is a power of $2$.

For the integration of a system of first order differential equations by
spectral methods one needs the primitive.
Now, let $F$ be a primitive of $f$
and $C_j\,,\;j=0,1,\ldots,N-1$ the coefficients of the spectral decomposition of $F$.
It turns out that a very simple relation relates the coefficients $C_j$ to the
$c_j$
\be
C_j=\frac{c_{j+1}-c_{j-1}}{2j} \;\;\;j>0\,,\quad{\rm with}\quad c_N=0\,.
\ee
As a primitive of $f$, the function $F$ has a single undetermined coefficient
$C_0$.
This coefficient has to be fixed
by the value of $F$ at some point in the interval.
Usually, the value of $C_0$ is fixed at one of the boundaries of
the interval $[-1,1]$. Indeed, we have
\bsub\
F(-1) & = & \frac{C_0}{2}+\sum_{k=1}^{N-1}\,C_k\,T_k(-1) =
\frac{C_0}{2}+\sum_{k=1}^{N-1}\,C_k\,(-1)^k\,,\\
F(1)  & = & \frac{C_0}{2}+\sum_{k=1}^{N-1}\,C_k\,T_k(1) =
\frac{C_0}{2}+\sum_{k=1}^{N-1}\,C_k\,.
\esub

\section*{Appendix 2}
In this Appendix we present the computation of the pertinent integrals (\ref{Energy-cut})
in detail.

We evaluate first the contribution to $d\tilde E(\beta)/d\beta$ from the far
region, ${\tilde E}^{'}_{\rm F}(\beta,r_{\rm m})$.
Using Eq.\ (\ref{Hfar}) to order $A_{\rm F}^2$ for $H_{\rm F}$ one finds that
it is given by
\beqn \label{exp-far1} %%
{\tilde E}^{'}_{\rm F}(\beta,r_{\rm m})&=&
\frac{1}{4\beta^2}\int_{x_{\rm m}}^{\infty}\,dx\,x^2(H_{\rm F}(x)^2-1)^2
 \nonumber \\
&\cong& \frac{A_{\rm F}^2}{\beta^2}\int_{x_{\rm m}}^{\infty}\,dx\,e^{-2x}
\l(1+A_{\rm F}\frac{e^{-x}}{x}+2A_{\rm F}e^{-x}f_1(x)\r) \nonumber\\
&=&\frac{A_{\rm F}^2}{\beta^2}\l[ -A_{\rm F}\ln x_{\rm m}-x_{\rm m}+\frac{1}{2}-A_{\rm F}\l(\gamma_{\rm E}+
\frac{\ln3}{4}+\frac{1}{2}\r)\r]+o(x_{\rm m})\,, \eeqn
where $x_{\rm m}=\beta r_{\rm m}$ ($r_{\rm m}$ denotes the matching point).
Using now $A_{\rm F}=-\beta-\beta^2(3\ln3)/4$, Eq.\ (\ref{exp-far1}) can be written as :
\be\label{exp-far11}
%\frac{d\tilde E_{\rm F}}{d\beta}=
\tilde E^{'}_{\rm F}(\beta,x_{\rm m})=
\beta \ln x_{\rm m} -x_{\rm m}+\frac{1}{2}+\beta\l(\frac{1}{2}+ \gamma_{\rm E}+\ln3\r)+o(x_{\rm m})\,.
\ee
To compute the monopole mass to order $\beta^2$ in the near region, $0<r<\Lambda$,
$H_{\rm N}$  should be identified with the BPS solution,
\be\label{HPS}
H_{\rm N}(r)=H_0(r) = \coth\,r-\frac{1}{r}\,,
\ee
and the contribution to $d\tilde E(\beta)/d\beta$ from the near
region is determined by
\beqn\label{dEdbetaNR}
  \int_{0}^{r_{\rm m}}dr\,r^2(H_0^2-1)^2&=&
  \left[4\,r+\frac{3}{r}-\frac{13}{3}\coth(r)+
   \frac{r(5\sinh(r)-r\,\cosh(r))}{3\,\sinh^3(r)}\right]^{r=r_{\rm m}}_{r=0}\nonumber \\
  && -\frac{2}{3}I_1(r_{\rm m}) -8I_2(r_{\rm m})\,,
\eeqn
with
$$
I_1(r_{\rm m})=\int_{0}^{r_{\rm m}}\,dr\frac{r^2}{\sinh^2(r)}\qquad
I_2(r_{\rm m})=\int_{0}^{r_{\rm m}}\,\frac{dr}{r}\left(\frac{1}{2}-\frac{1}{2r}+
\frac{1}{e^{2r}-1}\right)\,.
$$
The integral $I_1(r_{\rm m})$ is not difficult to compute:
\be\label{intI1}
I_1(r_{\rm m})=I_1(\infty)+o(1/r_{\rm m})=\frac{\pi^2}{6}+o(1/r_{\rm m})\,.
\ee
 In order to compute the last integral, $I_2(r_{\rm m})$, we consider the following identity
\be\label{limpto0}
I_2(r_{\rm m})=\lim_{p\to 0}
\bigg\{\int_{0}^{\infty}\,dx\,F(p,x)-\int_{2r_{\rm m}}^{\infty}\,dx\,F(p,x)\bigg\}\,,
\ee
with
\be
F(p,x) = \frac{e^{-px}}{x}\left(\frac{1}{2}
-\frac{1}{x}+\frac{1}{e^{x}-1}\right)\,.
\ee
Now, the first integral of the r.h.s. of (\ref{limpto0}) to order $O(p\ln p)$ is given by
\beqn \label{intp1}
\int_{0}^{\infty}\,dx\,F(p,x) =  \ln\Gamma(p)+\frac{1}{2}(\ln\,p-\ln\,2\pi)+p(1-\ln\,p)
%\nonumber\\
 \xrightarrow{p\to 0} -\frac{1}{2}(\ln\,p+\ln\,2\pi)\,,
\eeqn
while the second integral in Eq.\ (\ref{limpto0}) yields to order $r_{\rm m}^{-1}$
\beqn\label{intp2}
\int_{2r_{\rm m}}^{\infty}\,dx\,F(p,x) = -\frac{1}{2}\mbox{Ei}(-2pr_{\rm m})+O(r_{\rm m}^{-1})
\xrightarrow{p\to 0} -\frac{1}{2}(\gamma_E+\ln(2p\,r_{\rm m})+O(pr_{\rm m}))\,.\
\eeqn
Therefore
\be
I_2(r_{\rm m})=\frac{1}{2}\l(\ln r_{\rm m}+\gamma_{\rm E}-\ln\pi\r) +o(1/r_{\rm m})\,.
\ee
Putting now the above formulae together
leads to the result (up to $o(x_{\rm m})$):
\be\label{exp-near}
{\tilde E}^{'}_{\rm N}(\beta,r_{\rm m})
 =\frac{\beta}{4}\l(4r_{\rm m}-4\ln r_{\rm m}-C\r)
 =x_{\rm m}-\beta\ln\frac{x_{\rm m}}{\beta}-
 \beta\l(\gamma_{\rm E}+\frac{13}{12}+\frac{\pi^2}{36}-\ln\pi\r)\,.
\ee

Combining the contributions from the far (\ref{exp-far11}) and from the
near (\ref{exp-near}) regions, one finally obtains
\be\label{exp4}
\frac{d\tilde E(\beta)}{d\beta} =
\frac{1}{2}+\beta\ln\,\beta+\left(\frac{1}{2}
+\ln(3\pi)-\frac{13}{12}-\frac{\pi^2}{36}\right)\beta+o(\beta)\,.
\ee
Note: Equation (6.3) of Ref.\ \cite{Gardner82} corresponding to our Eq.\
(\ref{exp-near}) contains an important error. In \cite{Gardner82}
the value of the constant, $C$, is given (only numerically) as $C=0.9667$, whereas the
above analytical result yields $C=3.1599\ldots$. We have also verified
the correctness of $C=3.1599$ by a direct numerical calculation.

%%%%%%%%%%%%%%%%%%%%%%%%%%%%%%%%%%%%%%%%%%%%%%%%%%%%%%%%%%%%%%%%%%%%%%%%%%%%%%%%%%%%%%%

%%%%%%%%%%%%%%%%%%%%%%%%%%%%%%%%%%%%%%%%%%%%%%%%%%%%%%%%%%%%%%%%%%%%%%%%%%%%%%%%%%%%%%%

\end{document}